\documentclass[aps,prl,twocolumn,floatfix]{revtex4}
\usepackage{graphicx}
\usepackage{verbatim}

\begin{document}
\title{Cs in high oxidation states and as a p-block element}
\author{Mao-sheng Miao}
\affiliation{Materials Research Laboratory, University of California, Santa Barbara, CA 93106-5050, USA}
\affiliation{Beijing Computational Science Research Center, Beijing 10084, P. R. China}

\begin{abstract}
The major chemical feature of an element is the number of electrons available for forming chemical bonds. A doctrine rooted in the atomic shell model states that the atoms will maintain a complete inner shell while interacting with other atoms. Therefore, group IA elements, for example, are invariably  stable in the +1 charge state because the p electrons of their inner shells do not react with other chemical species. This general rule governs our understanding of the structures and reactions of matter and has never been challenged. In this work, we show for the first time that while mixing with fluorine under pressure, Cs atoms will share their 5p electrons and become oxidized to a higher charge state. The formal oxidation state can be as high as +5 within the pressure range of our study ($<$200 GPa) and stable Cs$^{2+}$ and Cs$^{3+}$ compounds can form at lower pressures. While sharing its 5p electrons, Cs behaves like a p-block element forming compounds with molecular, covalent, ionic and metallic features. Considering the pressure range required for the CsF$_{n}$ compounds, the inner shell electrons in other group IA and IIA elements may also bond with F or other chemical species under higher pressure. 
\end{abstract}

\maketitle

One of the important discoveries of the last century is the revelation of the atomic shell structure and the understanding of the periodic properties of elements at the quantum mechanics level.\cite{quantum-Laudau,Pauling-book} All elements except H and He possess completely filled inner shells, and all but noble gas elements possess a partially filled outer shell. The chemical properties of the atoms are determined by the electrons in the outermost shell; hence, these electrons are called valence electrons. Conversely, the electrons in the complete inner shells are not involved in interatomic bonding and are called core electrons. 

Based on the shell model, atoms form chemical bonds with other atoms by losing, gaining or sharing valence electrons, while all core electrons remain inert.\cite{Pauling-book,Murrel-book} The group I elements, for example, are invariably stable in a +1 charge state and usually form ionic compounds with other elements. The reactivity of a complete outmost shell was first demonstrated for noble gas elements by Neil Bartlett, who discovered in 1962 that Xe can react with F to form Xe$^{+}$[PtF$_{6}]^{-}$.\cite{Bartlett:1962} Since then, many noble gas compounds have been found.\cite{Agron:1963,Siegel:1963,Rundle:1963,Malm:1963,Liao:1998,Krouse:2007,Khriachtchev:2000,Grochala:2007,Dmochowski:2009,Brock:2011}  However, the inertness of the core electrons of all elements has not yet been seriously challenged. Several earlier attempts were made under ambient pressures,\cite{Bode:1950,Moock:1989} but these were later proven false.\cite{Asprey:1961,Jehoulet:1991} 

In this work, we demonstrate that the 5p electrons in the inner shell of a Cs atom can become reactive under high pressure. As a result, Cs atoms can be oxidized beyond the +1 oxidation state to form a series of fluorides CsF$_{n}$ (n=2 - 5) under pressures ranging from 10 to 200 GPa. Our approach is based on accurate first-principles calculations, which have been successfully used in numerous predictions regarding novel compounds and structures over the past few decades.\cite{Johnson:2000,Belonoshko:2003,Oganov:2009,Ma:2009} To search for stable CsF$_{n}$ structures under pressure, we employ a non-biased automatic structure search method based on the particle swarm optimization algorithm to search for stable structures across the entire energy surface.\cite{Wang:2012,Wang:2010}. The PSO structure predictions were performed by use of CALYPSO (crystal structure analysis by particle swarm optimization) for unit cells containing up to 4 CsF$_{n}$ units. In addition, more than 100 AB$_{n}$ structures in ICSD  (Inorganic Crystal Structure Database) were also included. Pressures of 0, 10, 50, 100, 150 and 200 GPa were chosen for our study. 

The underlying {\it ab initio} structural relaxations and the electronic band-structure calculations were performed within the framework of density functional method (DFT) as implemented by the VASP (Vienna Ab initio Simulation Package) code.\cite{Kresse:1996} The generalized gradient approximation (GGA) within the framework of Perdew-Burke-Ernzerhof (PBE)\cite{Perdew:1996} was used for the exchange-correlation functional, and the projector augmented wave (PAW) pseudopotentials\cite{Blochl:1994} were used to describe the ionic potentials. In the PAW potential for Cs, the 5s, 5p and 6s orbitals were included in the valence. The accuracy of the PAW potentials were tested and compared with a full-potential method. The cut-off energy for the expansion of the wave function into plane waves was set at 1200 eV, and Monkhorst-Pack $k$ meshes were chosen to ensure that all enthalpy calculations converged to better than 1 meV/atom.

\begin{figure}
\includegraphics[width=8.5cm]{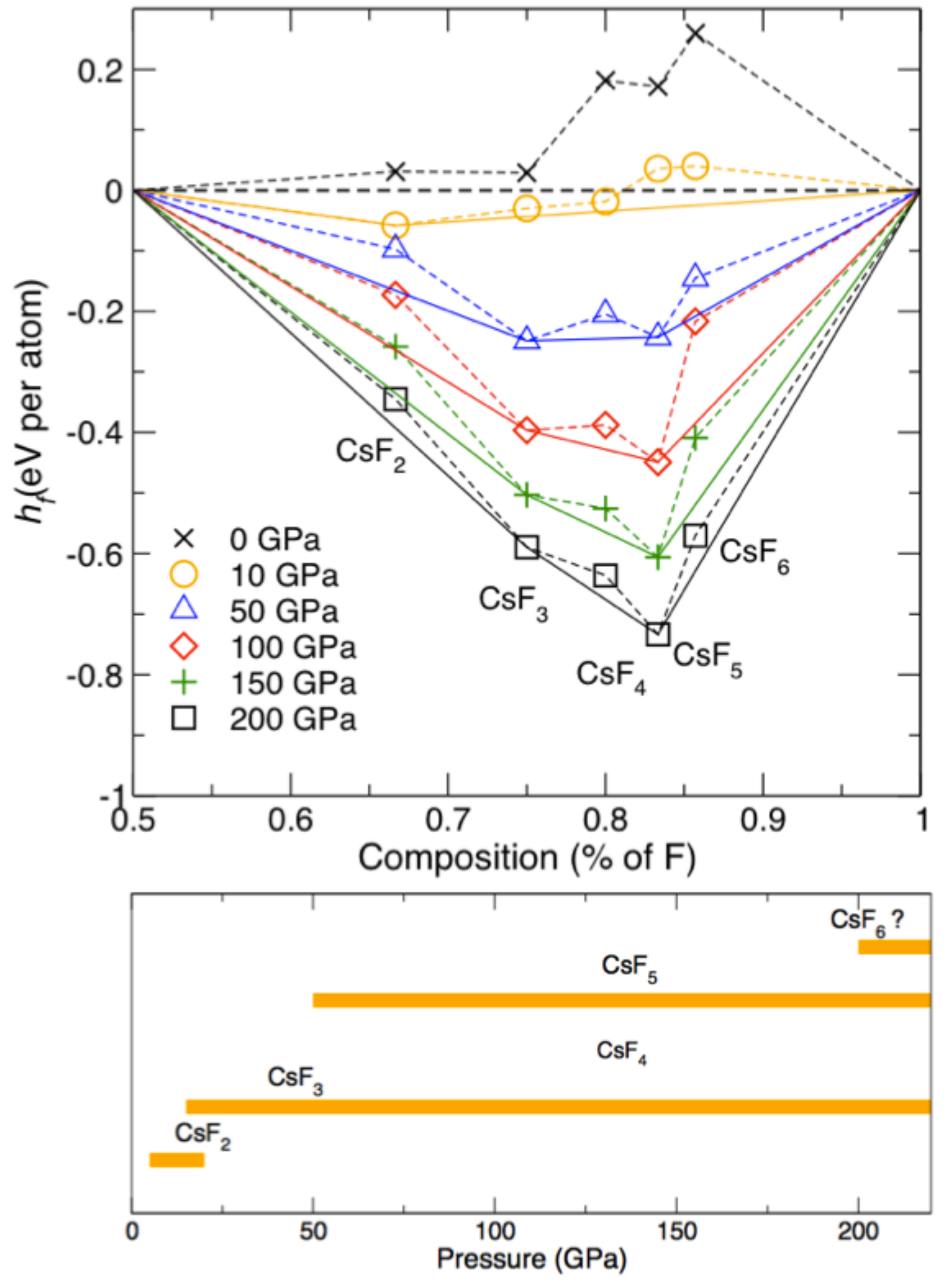}
\caption{\footnotesize (color online) {\bf (a)} The enthalpies of formation of CsF$_{n}$ under a series of pressures. The dotted lines connect the data points, and the solid lines denote the convex hull. {\bf (b)} The stable pressure range for CsF$_{n}$. 
\label{convex} }
\end{figure}

For the most stable structures at each pressure, the enthalpy of formation per atom is calculated using the following formula: $h^{f}(\textrm{CsF}_{n})=(H(\textrm{CsF}_{n}) - H(\textrm{CsF}) - (n-1)H(\textrm{F}_{2})/2)/(n+1)$.\cite{Feng:2008,Peng:2012} We choose to use the enthalpy of CsF instead of Cs (see the above formula) because of the substantial stability of CsF throughout the studied pressure range. Thus, changing Cs to CsF does not change the convexity of the compounds with higher F compositions, although the absolute value of $h^{f}$ does change.\cite{Peng:2012}

As shown in Fig. \ref{convex}, other than CsF, all CsF$_{n}$ compounds are unstable at ambient pressure, which is consistent with the common knowledge that Cs is only stable in the +1 state. However, CsF$_{2}$ becomes stable at a pressure of 10 GPa. As the pressure increases, CsF$_{n}$ compounds with higher F composition are formed. As shown in Fig. \ref{convex}(b), CsF$_{2}$ is only stable in a pressure range from 5 to 20 GPa. At higher pressures, the CsF$_{2}$ decomposes into CsF and CsF$_{3}$. The CsF$_{3}$ and CsF$_{5}$ become stable at pressures of 15 GPa and 50 GPa, respectively, and remain stable above 200 GPa. Our calculations show that CsF$_{4}$ and CsF$_{6}$ are unstable at pressures below 200 GPa, although the trend indicates that CsF$_{6}$ may become stable at higher pressures. The structures are all found to be dynamically stable within their stable pressure range. 

The structures of the CsF$_{n}$ compounds reveal that Cs can behave like a p-block element. While sharing its 5p electrons, Cs can covalently bond with F to form CsF$_{n}$ molecules. The corresponding CsF$_{n}$ compounds may have mixed molecular, covalent, ionic and possibly metallic features. 

\begin{figure}[tbp]
\includegraphics[width=8.5cm]{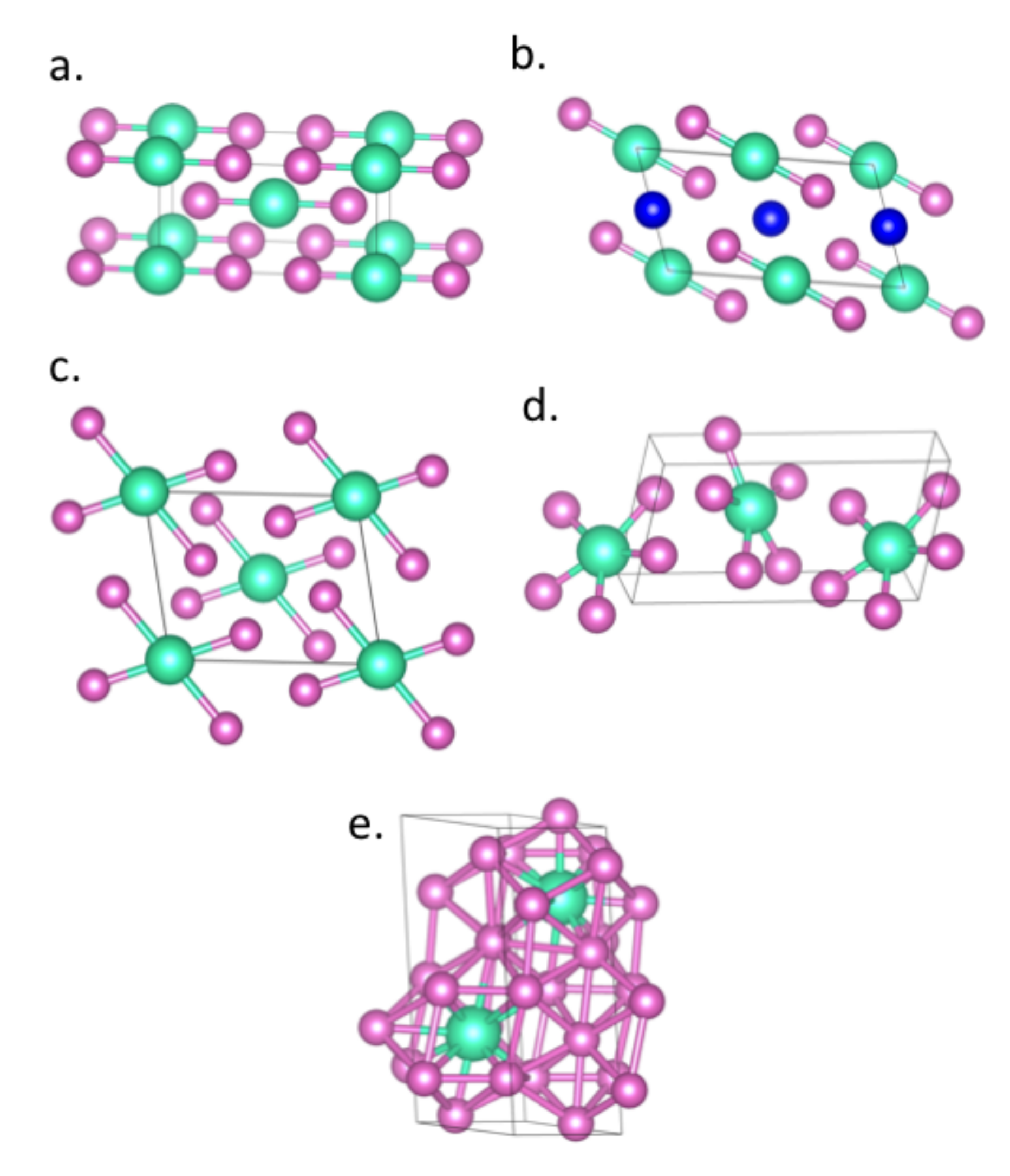}
\caption{\footnotesize (color online) The most stable structures of CsF$_{n}$ at selected pressures. {\bf (a)}: CsF$_{2}$ at 20 GPa in an I4/mmm structure; {\bf (b)}: CsF$_{3}$ at 50 GPa in a C2/m structure; {\bf (c)}: CsF$_{4}$ at 100 GPa in a C2/m structure; {\bf (d)}: CsF$_{5}$ at 100 GPa in an Fdd2 structure; {\bf (e)}: CsF$_{6}$ at 200 GPa in a P$_{1}$ structure. The green balls and the pink balls represent the Cs and the F atoms, respectively. The blue balls in {\bf (b)} represents the isolated F$^{-}$ ions in CsF$_{3}$. 
\label{structures} }
\end{figure}

As shown in Fig. \ref{structures}(a), the CsF$_{2}$ is stable in a structure with I4/mmm symmetry, similar to XeF$_{2}$.\cite{Agron:1963,Rundle:1963,Liao:1998} Linear CsF$_{2}$ molecules can be identified in this structure with a Cs-F bond length of 2.358 \AA~ at 20 GPa. This bond length is significantly larger than the Xe-F bond length of 1.999 \AA~, indicating a weaker Cs-F bond in CsF$_{2}$. Furthermore, the shortest F-F distances in CsF$_{2}$ are 2.215 \AA, which are much larger than the F-F bond length, suggesting that no covalent bond forms between the F atoms. The CsF$_{3}$ that is structured in a C2/m space group throughout the pressure range exhibits a distinctive molecular structure [Fig. \ref{structures}(b)]. Instead of forming a molecular crystal, the CsF$_{3}$ forms a [CsF$_{2}$]$^{+}$F$^{-}$ complex. At 100 GPa, the Cs-F bond length is 2.015 \AA~ close to that of Xe-F, whereas the next nearest Cs-F distance is 2.571 \AA. The shortest F-F distance is 2.227 \AA. CsF$_{4}$ is unstable. Some of its structures [Fig. \ref{structures}(c)] consist of CsF$_{4}$ molecules similar to XeF$_{4}$.\cite{Siegel:1963,Liao:1998} 

CsF$_{5}$ is stable in a structure with Fdd2 symmetry consisting of pentagonal planar molecules similar to XeF$_{5}^{-}$ [Fig. \ref{structures}(d)].\cite{Christe:1991} At 150 GPa, the 5 Cs-F bonds in the CsF$_{5}$ molecules have lengths of 1.886 \AA~, 1.899 \AA~ and 1.957 \AA. The distance to the next neighbouring F atoms is 2.367 \AA. The shortest F-F distance is 2.050 \AA. The P$_{1}$ structure of CsF$_{6}$ has the lowest energy. It consists of Cs atoms with 12 neighbouring F atoms that form a cage-like structure encircling the Cs atom and stack to form a crystal[Fig. \ref{structures}(e)]. 
\begin{figure}[tbh]
\hspace{1cm}
\includegraphics[width=9.cm]{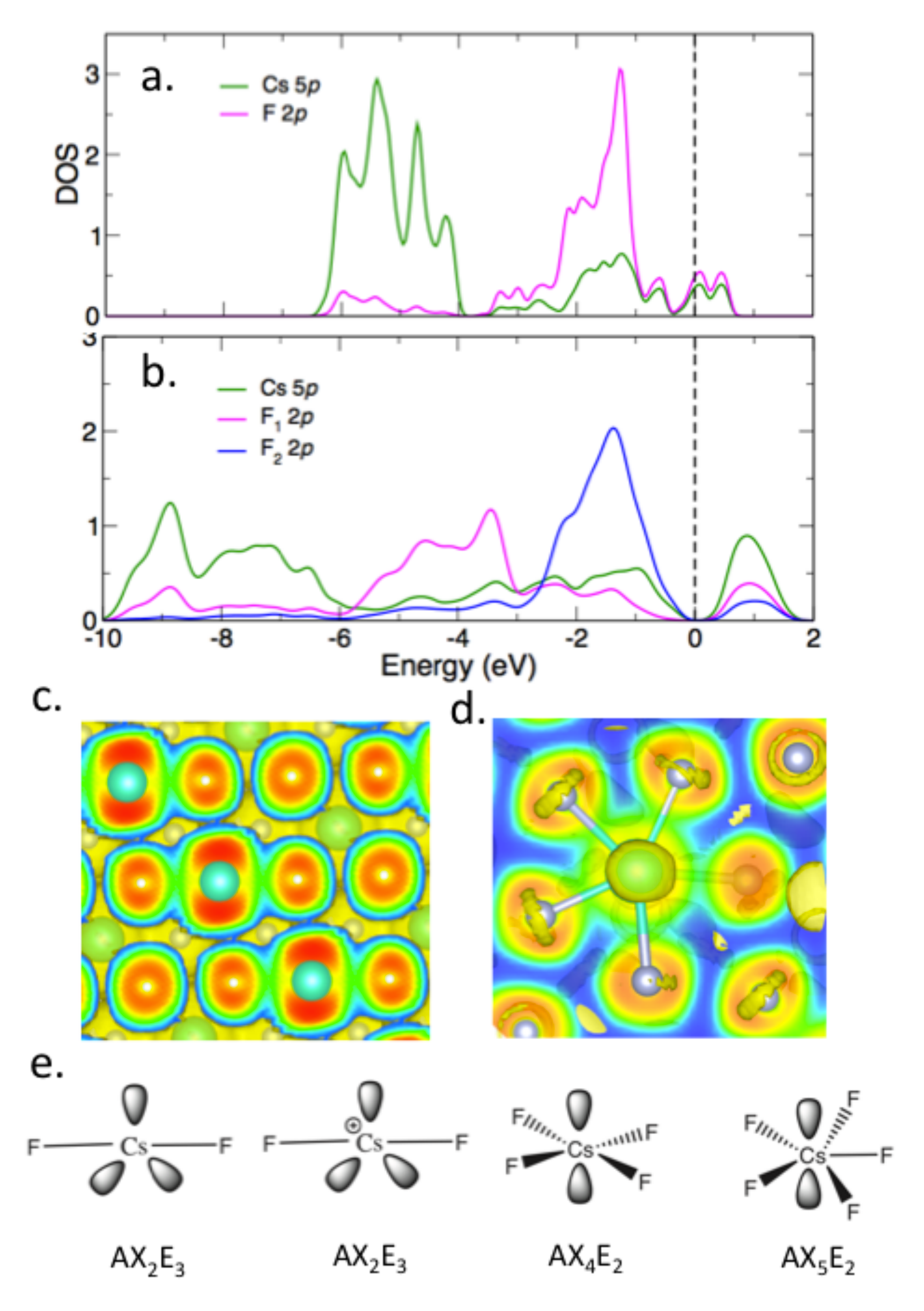}
\caption{\footnotesize (color online) {\bf (a)} and {\bf (b)}: the calculated projected density of states (PDOS) for CsF$_{2}$ at 20 GPa and and CsF$_{3}$ at 100 GPa, respectively. The green lines represent the 5p state of the Cs atoms, and the pink lines represent the 2p state of the F atoms bonded with Cs. The blue line in (b) represents the 2p state of the non-bonding F atom. {\bf (c)} and {\bf (d)} The calculated electron localization functions (ELF) of CsF$_{3}$ at 50 GPa and CsF$_{5}$ at 100 GPa. {\bf (e)} The bonding feature of CsF$_{n}$ (n=2-5), given in valence-shell electron-pair repulsion (VSEPR) notation. 
\label{bonding} }
\end{figure}

\begin{figure}[tbh]
\includegraphics[width=8.5cm]{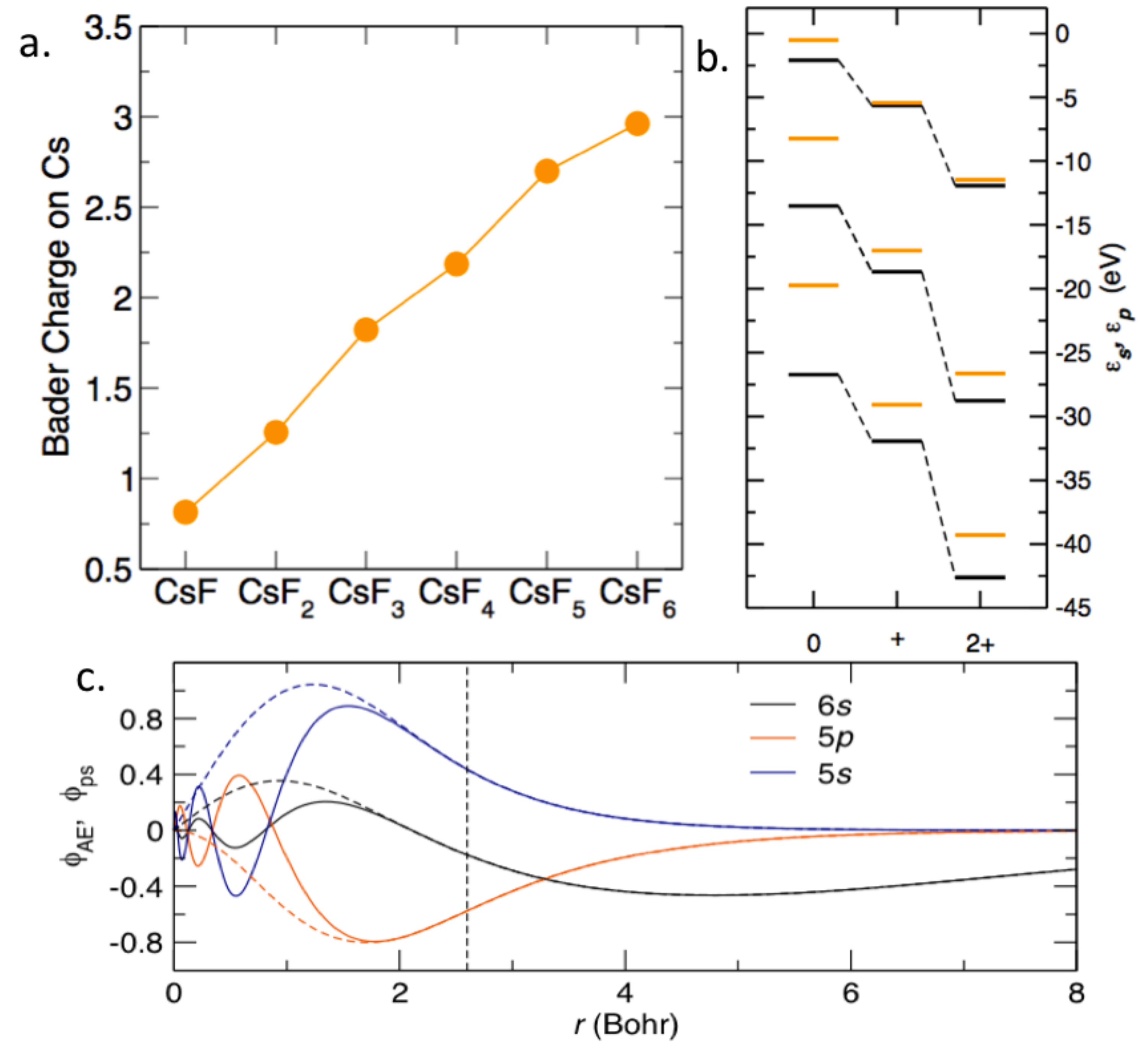}
\caption{\footnotesize (color online) {\bf (a)} The calculated Bader charge of Cs in CsF$_{n}$ at 100 GPa. {\bf (b)} The 6s, 5p and 5s energy levels of the Cs atom in neutral, +1 and +2 charge states. A comparison of the same levels for Xe$^{0}$, Xe$^{+}$ and Xe$^{2+}$ are shown by the orange bars. {\bf (c)} The radial wave function of the Cs 6s (black), 5p (orange) and 5s (blue) states. The solid and dashed lines indicate the all-electron and pseudo-wave functions, respectively.
\label{bader} }
\end{figure}

The electronic structures of CsF$_{n}$ (n$\geq$2) also reveal that the 5p electrons of Cs involve in forming chemical bonds with neighboring F atoms. The calculated PDOS shows large overlap between the Cs 5p states and the F 2p states [Fig. \ref{bonding}(a) and (b)]. More remarkably, there are large 5p components in the conducting states, indicating the depletion of 5p electrons in Cs atoms in these compounds. Furthermore, CsF$_{3}$ and CsF$_{5}$ appear to have gaps, whereas the CsF$_{2}$ are metallic. The charge distribution and the electron localization function (ELF) show multiple covalent bonding between Cs atoms and their neighboring F atoms.[Fig. \ref{bonding}(c) and (d)]  The bonding features of CsF$_{n}$ molecules are analogous to the isoelectronic XeF$_{n}$ molecules. 

The molecular orbitals and therefore the bonding features of CsF$_{2}$ are analogous to XeF$_{2}$ that is in  AX$_{2}$E$_{3}$ structure [Fig. \ref{bonding}(c) and (d)]. The  extra electron of CsF$_{2}$ fills the Cs-F p$_{z}$-p$_{z}$ $\sigma$ anti-bonding state, leading to weakened Cs-F bonds and a metallic character in the system. The [CsF$_{2}$]$^{+}$ molecular ions in CF$_{3}$ is isoelectronic with a XeF$_{2}$ molecule. A gap exists between the $\pi$ anti-bonding and the $\sigma$ anti-bonding states, and CsF$_{3}$ maintains a strong molecular character and stability over a large pressure range. 

CsF$_{5}$ molecules are pentagonal planar, corresponding to a structure of AX$_{5}$E$_{2}$ [Fig. \ref{bonding}(c) and (d)]. The structure differs from that of most other AB$_{5}$ molecules, such as BrF$_{5}$, in that it adopts a square pyramidal geometry. A similar pentagonal planar structure is found in [XeF$_{5}$]$^{-}$,\cite{Christe:1991} a species that is isoelectronic with the CsF$_{5}$ molecule. 

The atomic and electronic structural features indicate that the oxidation state of Cs in CsF$_{n}$ is +n. To further demonstrate this, we calculated the charges using Bader analysis for CsF$_{n}$ under a pressure of 100 GPa. As shown in Fig. \ref{bader}(a), the Bader charges increase almost linearly with an increasing number of F atoms in the chemical formula and are seemingly larger than +1 for CsF$_{n}$ ($n\ge$2) compounds. As expected, the Bader charge is notably smaller than the formal oxidation number, even for CsF. 

The potential for Cs to be oxidized to a high charge state can be revealed by the energy and the geometry of the atomic orbitals. Figure \ref{bader} compares the 6s, 5p and 5s energy levels of Xe and Cs at various charge states, calculated using DFT for a single atom. The results indicate that although the 5p state of Cs$^{0}$ is 5.28 eV lower than that of Xe$^{0}$, the difference is reduced to only 1.66 eV for the +1 charge states and to 2.14 eV for the +2 charge states. Because Xe can be oxidized up to a +8 charge state, one might expect that Cs could also be oxidized beyond the +1 state. The Cs 5p orbital peaks at 0.98 \AA~ and has a large component outside the Cs$^{+}$ radius of 1.81 \AA[Fig. \ref{bader}(c)]. Under pressure, the Cs-F distance may become much smaller than the sum of the Cs$^{+}$ and F$^{-}$ radii (~3.0 \AA). Furthermore, the Cs 5p bands will be significantly broadened by pressure, which will elevate the energy of some 5p states and prompt a sharing of the Cs 5p electron with the F atom.

In summary, we demonstrate using first principles calculations that the Cs atoms will inevitably be oxidized to high charge states by F atoms under pressures that are accessible by current high pressure technique. A series of compound of CsF$_{n}$ will be stabilized, in which Cs has formal oxidation number of +n. In these compounds, the 5p electrons of Cs are involved in forming chemical bonds with neighboring F atoms. Cs behaves like a p block element and forms CsF$_{n}$ molecules analogous to XeF$_{n}$. Based on our results, the long time dogma that the complete inner shell of an element is inert to chemical reactions, an important aspect of the atomic shell model, is no longer true. Considering the pressure range for CsF$_{n}$ compounds, the inner shell electrons in other group IA and IIA elements may also involve in bonding activities with F or other chemical species under higher pressure. 
\\

\normalsize
\noindent{\bf Acknowledgement} The author thanks Prof. Roald Hoffmann in Cornell University and Prof. Ram Seshadri in UCSB for inspiring discussions and constructive suggestions. This work is supported by the MRSEC program (NSF- DMR1121053) and the ConvEne-IGERT Program (NSF-DGE 0801627). The calculations made use of CNSI computing clusters and NSF-funded XSEDE resources. 
\\

\bibliography{references}

\hspace{0.1in}

\end{document}